\documentclass[12pt]{iopart}
\usepackage{overpic}
\usepackage{subfig}
\usepackage{iopams}
\usepackage{graphicx}
\usepackage{url}

\makeatletter
\renewcommand\section{\@startsection
  {section}{1}{0mm}
  {\baselineskip} 
  {0.5em}         
  {\normalfont\normalsize\bfseries}}
\renewcommand\subsection{\@startsection
  {subsection}{2}{0mm}
  {\baselineskip}
  {0.5em}
  {\normalfont\normalsize\bfseries}}
\renewcommand\subsubsection{\@startsection
  {subsubsection}{3}{0mm}
  {\baselineskip}
  {0.5em}
  {\normalfont\normalsize\bfseries}}
\makeatother

\begin{document}

\title[Generative Lagrangian Data Assimilation]%
      {Generative Lagrangian data assimilation for ocean dynamics under extreme sparsity}

\author{Niloofar Asefi$^{2}$, Leonard Lupin-Jimenez$^{1}$, Tianning Wu$^{3}$, Ruoying He$^{3}$, and Ashesh Chattopadhyay$^{1}$}

\address{$^1$Department of Applied Mathematics, University of California, Santa Cruz, CA 95064, USA}
\address{$^2$Department of Electrical and Computer Engineering, University of California, Santa Cruz, CA 95064, USA}
\address{$^3$Department of Marine, Earth, and Atmospheric Sciences, North Carolina State University, Raleigh, NC 27695, USA}
\ead{aschatto@ucsc.edu}

\begin{abstract}
Reconstructing ocean dynamics from observational data is fundamentally limited by the sparse, irregular, and Lagrangian nature of spatial sampling, particularly in subsurface and remote regions. This sparsity poses significant challenges for forecasting key phenomena such as eddy shedding and rogue waves. Traditional data assimilation methods and deep learning models often struggle to recover mesoscale turbulence under such constraints. We leverage a deep learning framework that combines neural operators with denoising diffusion probabilistic models (DDPMs) to reconstruct high-resolution ocean states from extremely sparse Lagrangian observations. By conditioning the generative model on neural operator outputs, the framework accurately captures small-scale, high-wavenumber dynamics even at $99\%$ sparsity (for synthetic data) and $99.9\%$ sparsity (for real satellite observations). We validate our method on benchmark systems, synthetic float observations, and real satellite data, demonstrating robust performance under severe spatial sampling limitations as compared to other deep learning baselines.

\end{abstract}

%
\vspace{2pc}
\noindent{\it Keywords}: Lagrangian, sparse observations, conditional diffusion, data assimilation, ocean dynamics 
%
%
%
%

\section{Introduction}

Forecasting multi-scale chaotic dynamical systems, e.g., the Earth's ocean, requires a dynamical model of the system's equations and accurate initial conditions. These initial conditions are obtained from satellite altimeters, ocean buoys, drifters, etc.~\cite{moore2019synthesis,dickey2003emerging,kubota2002japanese}. All of these measuring devices that observe the different variables of the ocean inherently provide noisy and sparse observations. On top of that, the location at which we observe the ocean moves with these measuring devices. Such observations are referred to as Lagrangian observations~\cite{molcard2005lagrangian,salman2006method,kuznetsov2003method,apte2008bayesian}. The sparsity and Lagrangian nature of these observations make it challenging to reconstruct the turbulent flow fields accurately, especially for the small-scale high-wavenumber dynamics~\cite{bennett2005inverse,molcard2003assimilation,oke2009comparison,schulz2010statistical}. The accuracy of these initial conditions affects the quality and time horizon of the forecasts, especially for extreme events~\cite{martin2007data,lermusiaux2006uncertainty,smith2014new,edwards2015regional} which have a major societal impact.

Traditional approaches to assimilating sparse Lagrangian observations rely on Bayesian ensemble-based methods such as the ensemble Kalman filter (EnKF)~\cite{evensen2003ensemble} and ensemble optimal interpolation~\cite{oke2008representation,oke2010ocean,xie2010ensemble}. These methods generate ensembles of background states using a dynamical model to estimate the background error covariance. However, accurately capturing high-dimensional covariances typically requires a large number of ensemble members, which is often computationally prohibitive. In practice, the ensemble size is severely limited, leading to rank-deficient covariance matrices. To mitigate this issue, practitioners employ heuristic localization techniques~\cite{oke2007impacts} to constrain spurious long-range correlations and improve assimilation performance under limited ensemble sizes.

With the rise of data-driven models for atmospheric~\cite{pathak2022fourcastnet,lam2022graphcast,guan2024lucie,watt2023ace,watt2025ace2} and oceanic dynamics~\cite{chattopadhyay2024oceannet,lowe2025long,gray2025long,dheeshjith2025samudra}, trained predominantly on reanalysis datasets, the importance of accurate initial conditions has become increasingly critical—particularly as these models are being adapted for operational use with direct initialization from observations~\cite{chen2023towards,mcnally2024data,allen2025end}. Most efforts toward operational deployment rely on sparse observations from satellite remote sensing and weather stations. Unlike traditional numerical models, data-driven models inherently lack explicit physical constraints, making them more sensitive to initialization errors. As a result, inaccuracies in initial conditions can degrade forecast skill rapidly. Although data-driven analogues of ensemble Kalman filtering and variational assimilation have shown promise in geophysical settings~\cite{chattopadhyay2022deep,chattopadhyay2021towards,maulik2022efficient}, these methods still depend on a capable background model—one that can either produce a sufficiently large and diverse ensemble of plausible background states (for EnKF) or enable gradient-based optimization through adjoint computations for variational approaches.

In this study, we address the challenge of generating accurate and physically consistent initial conditions for surface ocean dynamics from \textbf{extremely sparse} ($99\%$ and $99.9\%$) \textbf{Lagrangian} observations in the \textbf{absence} of a background model. To achieve this, we propose a neural operator-conditioned generative diffusion model. While related work by Scott et al.~\cite{martin2024genda} has explored this problem under nominal sparsity and fixed observational locations in the open ocean, our contributions focus on the more complex setting of regional ocean dynamics, characterized by intricate bathymetry, coastal boundaries, and highly sparse, moving observations.

Our approach draws inspiration from Oommen et al.~\cite{oommen2024integrating} and combines the predictive capabilities of a Fourier Neural Operator (FNO) and a UNET  
with a conditional generative diffusion model, denoising diffusion probabilistic model (DDPM), enabling the reconstruction of high-resolution ocean states from sparse Lagrangian observations. This integrated FNO+DDPM or UNET+DDPM architecture demonstrates significantly improved reconstruction accuracy compared to conventional machine learning-based methods.

Moreover, we show that under such extreme sparsity, standard statistical evaluation metrics such as root mean squared error (RMSE), correlation coefficient (CC), and structural similarity index (SSIM) tend to emphasize large-scale structures and fail to capture the full quality of the reconstructed turbulent ocean fields due to spectral bias~\cite{chattopadhyay2023long}, wherein deep neural networks preferentially learn low-wavenumber dynamics while missing critical fine-scale dynamics. In contrast, our proposed model preserves both large- and small-scale features of the flow field, as demonstrated through detailed Fourier spectral analysis and other physics-informed diagnostics.

To validate our framework, we conduct a comprehensive evaluation on three systems: (a) synthetic sparse Lagrangian tracers from a canonical geophysical fluid dynamics benchmark: two-dimensional forced turbulence on a $\beta$-plane with multiple zonal jets~\cite{maltrud1991energy}, simulated at Reynolds number $Re = 10{,}000$; (b) synthetic sparse Lagrangian tracers from a high-resolution surface ocean reanalysis data over the Gulf of Mexico, obtained from the $\frac{1}{12}^{\circ}$ resolution GLORYS dataset~\cite{garric2018performance}; and (c) real-world satellite altimetry observations over the Gulf of Mexico~\cite{he2025advanced}, which were used to generate a novel high-resolution ($\frac{1}{25}^{\circ}$) reanalysis dataset, CNAPS.

\section{Methods}

\subsection{Systems}

In this section, we discuss the three different oceanic systems: a canonical high-resolution fluid flow simulation, regional high-resolution reanalysis data, and real satellite observations, for which we have validated our proposed model.  

\textbf{System 1:} We consider the 2D forced Kolmogorov flow on a $\beta$-plane at $Re = 10{,}000$. The governing equations for the flow are:

\begin{equation}
\frac{\partial \omega}{\partial t} + \mathbf{u} \cdot \nabla \omega + \beta v = \frac{1}{Re} \nabla^2 \omega + \sin(4x)+\sin(4y),
\label{eq:vorticity}
\end{equation}

\begin{equation}
\omega = \nabla^2 \psi,
\label{eq:poisson}
\end{equation}

\begin{equation}
\mathbf{u} = (-\partial_y \psi, \partial_x \psi),
\label{eq:velocity}
\end{equation}

where $\beta = 20$ is the Coriolis parameter, $\omega$ is the vorticity, $\mathbf{u}$ is the velocity vector related to the stream function, $\psi$ through Eq.~(\ref{eq:velocity}). The equations are forced with sinusoidal forcing with both zonal and meridional wavenumber as $4$. $Re$ is set to $10{,}000$. We perform a direct numerical simulation (DNS) on this system from a random initial condition until steady-state turbulence is achieved. We utilize a doubly periodic pseudo-spectral solver with a $\Delta t_{DNS}=5\times 10^{-3}$ and $256\times 256$ spatial grid. The non-zero value $\beta$ in this case leads to the generation of zonal jets in this simulation. 

 We then extract $7000$ temporal samples of $\omega(t)$ at every $20 \Delta t_{\mathrm{DNS}}$. To generate synthetic observations, we randomly remove $99\%$ of the spatial grid points for every sample. Each of these $7000$ sparse samples with $1\%$ of grid points retained randomly over the domain are analogous to Lagrangian observations on which our proposed models (see section~\ref{sec:models}) are trained with their corresponding full-resolution $\omega(t)$ as targets. An independent $100$ temporal $\omega(t)$ samples with $99\%$ sparsity are used to test the models' performance.

\textbf{System 2:} The second system that we consider is surface ocean dynamics, i.e., sea-surface height (SSH), sea-surface zonal velocity (SSU), and sea-surface meridional velocity (SSV) over the Gulf of Mexico from the GLORYS reanalysis dataset~\cite{garric2018performance} at $\frac{1}{12}^{\circ}$ horizontal resolution. This reanalysis data is obtained by integrating observations with the NEMO ocean model~\cite{storkey2010forecasting}. To simulate the synthetic Lagrangian observations, we utilize the same procedure outlined in \textbf{System 1}. We used a total of $7000$ temporal samples of SSH, SSU, and SSV from $1993$ to $2012$ for training. We report the performance of our proposed model on $50$ independent temporal samples from outside the training set. 

\textbf{System 3:} Finally, we validate our proposed framework on a third system using real satellite altimetry observations and a corresponding high-resolution reanalysis dataset, CNAPS~\cite{he2025advanced}, which was produced by assimilating these observations into the Regional Ocean Modeling System (ROMS). Unlike Systems 1 and 2, which rely on synthetically generated observations, this system leverages actual satellite-derived sea surface height (SSH) data that are inherently Lagrangian and extremely sparse, with approximately $99.9\%$ spatial sparsity. In this setting, the model is trained to reconstruct full-resolution SSH fields from these sparse observations using the CNAPS reanalysis as the supervisory target (details provided in Section~\ref{sec:models}). The SSH observations are derived from multiple satellite altimetry missions starting from January 1, 1993. Due to the nadir-only measurement geometry of satellite tracks, the data exhibit strong spatial sparsity and highly non-uniform spatio-temporal sampling. We source these SSH data from the Copernicus Marine and Environment Monitoring Service (CMEMS), where they are processed through the Data Unification and Altimeter Combination System (DUACS). This processing pipeline includes inter-mission homogenization to a unified reference frame, as well as corrections for dynamic atmospheric effects, tidal signals, and long-wavelength errors. To match the reanalysis grid, the daily altimetry data are transformed into super-observations via spatial averaging within each CNAPS grid cell. To train the model, we have used a total of $4017$ samples from $1993$ to $2003$. We have validated the performance on $50$ independent samples from outside the training set. 

\subsection{Models}
\label{sec:models}
    \subsubsection{UNET}
    \vspace{0.5em}
    
Our baseline method is a UNET model, which is a convolutional neural network (CNN) and was introduced primarily for the application of biomedical image segmentation by Ronneberger et al.~\cite{ronneberger2015unet}. The UNET architecture comprises two main parts: the encoder (contracting path) and the decoder (expanding path). In our UNET architecture, the encoder comprises two convolutional blocks, each containing two convolutional layers, followed by max-pooling layers. Each convolution layer performs a convolution operation utilizing k×k filters, optionally employing padding to maintain spatial dimensions. The output of each convolutional block is further processed by a nonlinear activation function named the Rectified Linear Unit (ReLU). 
The bottleneck block connects the encoder and decoder. Our UNET has two convolutional layers utilizing 3×3 kernels, each followed by a ReLU activation function. It transforms 128 input channels into 256 output channels while maintaining spatial resolution. In our UNET model, the decoder consists of upsampling operations followed by convolutional layers and is structurally symmetric to the encoder. In our framework, upsampling is performed by transposed convolutions (or deconvolutions) using a kernel size of 2×2 and a stride of 2, which doubles the spatial resolution at each step. Following each upsampling operation, the relevant feature map from the encoder is concatenated with the decoder's feature map through skip connections, thus maintaining crucial spatial details. Each concatenated feature map is then passed through a convolutional block of two convolutional layers with 3×3 kernels, each followed by a ReLU activation function. Our decoder has two convolutional blocks that progressively enhance spatial resolution while refining feature representations. The final layer of our UNET model is a
1×1 convolution that reduces the feature map to a single output channel. We trained the UNET model on extremely sparse observations (e.g., partial SSH, SSU, and SSV) at time $t$ to predict the full-resolution ground truth at the same time $t$, by minimizing the L2 norm, i.e., the Mean Squared Error (MSE) loss.

    \subsubsection{FNO}
    \label{sec:FNO}
Neural operators are deep learning architectures designed to approximate mappings between infinite-dimensional function spaces. Formally, given an operator $\mathcal{G}$ that maps an input function $X$ to an output function $Y$ (i.e., $\mathcal{G}: X \rightarrow Y$), a neural operator seeks to learn this functional relationship and predict $y \in Y$ for a given $x \in X$. These models are increasingly used to learn solution operators of partial differential equations (PDEs) and other functional mappings that arise in dynamical systems. However, like conventional neural networks, they exhibit spectral bias—favoring low-wavenumber components during training—which can limit their ability to reconstruct fine-scale features~\cite{oommen2024diffusion}.

In this work, we utilize the \textit{Fourier Neural Operator} (FNO), a specific class of neural operator introduced by Li et al.~\cite{li2020fno}. FNOs operate in the spectral domain by applying a Fourier transform to the input data, truncating the higher frequency modes, and applying learned complex-valued weights to the retained spectral coefficients. The modified spectrum is then transformed back into the spatial domain via an inverse Fourier transform and added to the result of a pointwise linear transformation of the original input. Unlike traditional convolutional networks, FNOs capture long-range spatial dependencies efficiently and are resolution-agnostic, allowing them to generalize across varying input grid sizes without retraining on fixed-resolution data.

Our FNO architecture comprises four Fourier layers, each with a width of $20$. The input to the model is  $u_{\mathrm{partial}}(x, y, t)$, which are the sparse observations from either of \textbf{System 1}, \textbf{2}, or \textbf{3}. The model is trained using the Adam optimizer with a learning rate of 0.001 to minimize the $L^2$ loss between predicted and ground truth fields. 

\begin{equation}
\min_{\theta_{\mathrm{NO}}} \left\| u(x, y, t) - \mathcal{G}_{\theta_{\mathrm{NO}}} \left( u_{\mathrm{partial}}(x, y, t) \right) \right\|^2_2,
\label{eq:1}
\end{equation}

where $u(x, y, t)$ denotes the full-resolution target field, and $u_{\mathrm{partial}}(x, y, t)$ represents the sparse input observations. The operator $\mathcal{G}_{\theta_{\mathrm{NO}}}$, parameterized by $\theta_{\mathrm{NO}}$, denotes the FNO model that learns to approximate the functional mapping from partial observations to complete spatiotemporal fields.

\subsubsection{FNO-conditioned DDPM (FNO+DDPM)}

Following recent success in computer vision, we employ a Conditional Denoising Diffusion Probabilistic Model (DDPM) with a UNET-based denoiser conditioned on the output of the FNO model described in section~\ref{sec:FNO}, to reconstruct full-resolution ground truth fields from sparse observational data. The objective is to model the conditional distribution \( p_\theta(u \mid u_{\mathrm{partial}}) \), where \( u \) is the full-resolution state (e.g., sea surface height), and \( u_{\mathrm{partial}} \) is a sparsely observed input.

The key innovation is to condition the DDPM's denoiser, \( \epsilon_\theta(u_\tau, \tau) \), on the output of the FNO model, denoted \( \mathcal{G}_{\theta_{\mathrm{NO}}}(u_{\mathrm{partial}}) \). Here, \( u_0 \) denotes the original ground truth field at time \( t \), and \( u_\tau \) represents a noised version at diffusion timestep \( \tau \in \{1, \dots, T\} \). The model is trained on individual snapshots of the full-resolution field \( u_0 \), with diffusion applied over \( T = 1000 \) time steps.

\begin{enumerate}
    \item \textbf{Conditional Input to the Network:} \\
    Rather than learning \( \epsilon_\theta(u_\tau, \tau) \) directly, our model learns the denoiser:
    \[
    \epsilon_\theta(u_\tau, \tau, \mathcal{G}_{\theta_{\mathrm{NO}}}(u_{\mathrm{partial}})),
    \]
    where the FNO output is concatenated with the noisy input \( u_\tau \). The denoiser is trained to predict the added Gaussian noise at each diffusion step, using a UNET architecture.

    \item \textbf{Training Objective:} \\
        The training loss follows the simplified DDPM objective:
        \begin{equation}
        \mathcal{L}_{\mathrm{simple}}(\theta) := \mathbb{E}_{\tau, u, \epsilon} \left[ \left\| \epsilon - \epsilon_\theta\left( \sqrt{\bar{\alpha}_\tau} u + \sqrt{1 - \bar{\alpha}_\tau} \, \epsilon, \, \tau, \, \mathcal{G}_{\theta_{\mathrm{NO}}}(u_{\mathrm{partial}}) \right) \right\|_2^2 \right],
        \label{eq:loss_simple}
        \end{equation}
        where \( \epsilon \sim \mathcal{N}(0, \mathbf{I}) \) is sampled Gaussian noise.
        
    \item \textbf{Forward Diffusion Process:} \\
    The forward diffusion process adds noise to the full-resolution field as follows:
    \begin{equation}
    q(u_\tau \mid u_0) = \mathcal{N}\left( u_\tau;\, \sqrt{\bar{\alpha}_\tau} u_0,\, (1 - \bar{\alpha}_\tau) \mathbf{I} \right),
    \label{eq:forward_process}
    \end{equation}
    where \( \bar{\alpha}_\tau = \prod_{s=1}^{\tau} (1 - \beta_s) \) is the cumulative product of the noise schedule.

    \item \textbf{Reverse Process:} \\
    The FNO prediction provides both a conditioning prior and the initial guess for the reverse process. The initial noisy sample is generated as:
 
    \begin{equation}
        u_T = \sqrt{\bar{\alpha}_T} \, \mathcal{G}_{\theta_{\mathrm{NO}}}(u_{\mathrm{partial}})
        + \sqrt{1 - \bar{\alpha}_T} \cdot \epsilon,
        \quad \epsilon \sim \mathcal{N}(0, \mathbf{I}).
    \label{eq:init_reverse}
    \end{equation}
    
    At each diffusion step \( \tau \), the sample is updated using the learned denoiser:

    \begin{equation}
        u_{\tau - 1} 
        = \frac{1}{\sqrt{\alpha_\tau}} 
        \left( 
            u_\tau 
            - \frac{1 - \alpha_\tau}{\sqrt{1 - \bar{\alpha}_\tau}} 
            \cdot \epsilon_\theta
            \bigl(
                u_\tau, \tau, \mathcal{G}_{\theta_{\mathrm{NO}}}(u_{\mathrm{partial}})
            \bigr)
        \right) 
        + \sigma_\tau z.
    \label{eq:reverse_step}
    \end{equation}

    where \( \sigma_\tau = \sqrt{\beta_\tau} \) and \( z \sim \mathcal{N}(0, \mathbf{I}) \) if \( \tau > 1 \), otherwise \( z = 0 \).

    \item \textbf{Beta Schedule:} \\
    The noise variance \( \beta_\tau \) is defined using a linear schedule:
    \begin{equation}
        \beta_\tau 
        = \beta_{\mathrm{start}} 
        + \left( 
            \frac{\tau - 1}{T - 1} 
          \right)
          \bigl(
            \beta_{\mathrm{end}} - \beta_{\mathrm{start}}
          \bigr).
    \label{eq:beta_schedule}
    \end{equation}
    
    where \( \beta_{\mathrm{start}} \) and \( \beta_{\mathrm{end}} \) are the minimum and maximum noise levels used across \( T = 1000 \) diffusion steps.
\end{enumerate}

This setup enables the DDPM to generate physically consistent, full-resolution reconstructions by leveraging the coarse predictions from the FNO as a learned physics-informed prior. Each denoising step is guided by both the noisy input and the global spatial structure inferred by the FNO from sparse observations.

\begin{figure}
  \centering
  \includegraphics[width=\linewidth]{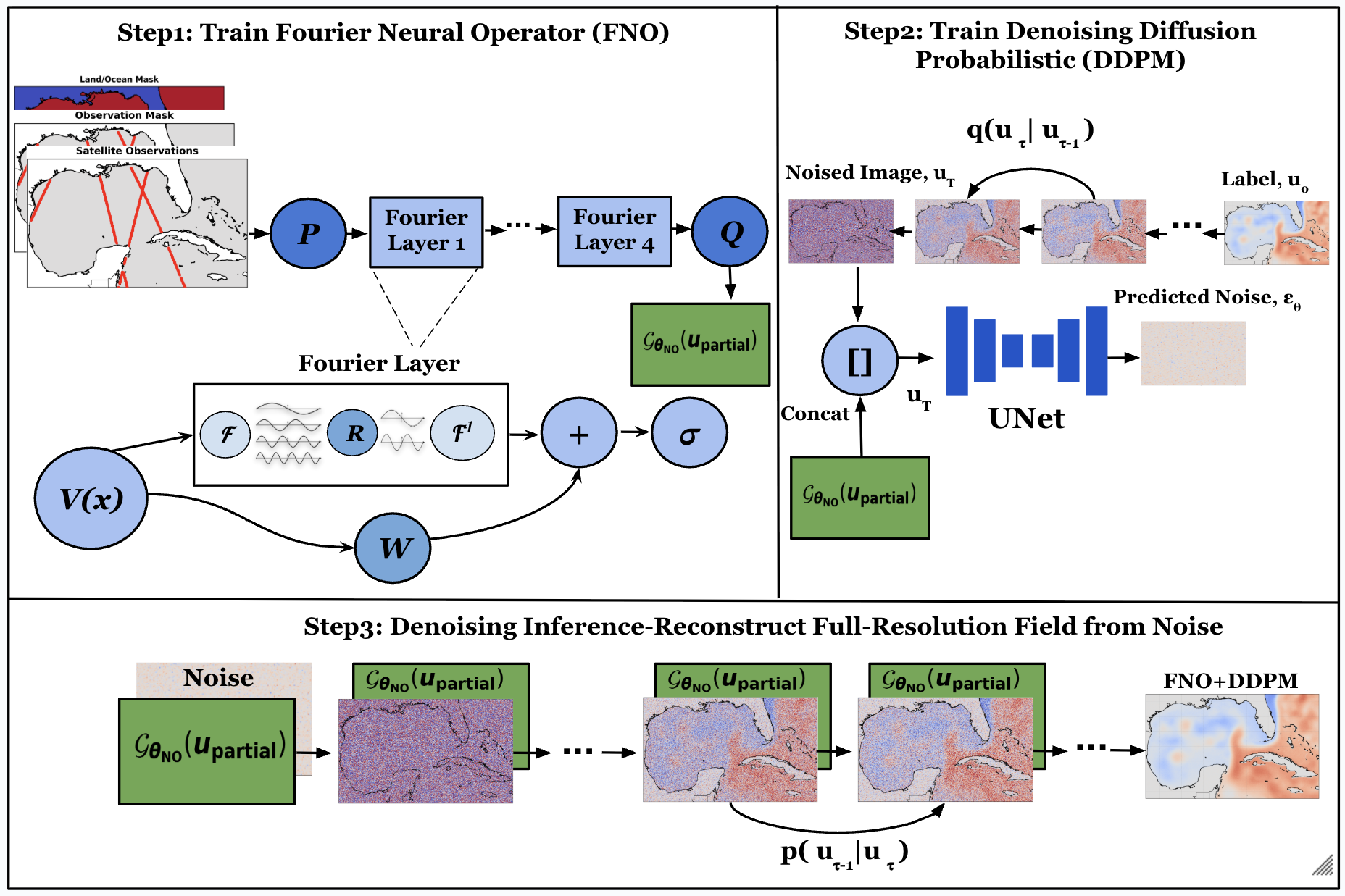}
  \caption{Schematic of the FNO-conditioned diffusion framework (FNO+DDPM). The Fourier Neural Operator (FNO) is first trained to predict full-resolution ocean states from highly sparse observational inputs. These coarse reconstructions are then used as conditioning inputs to a Denoising Diffusion Probabilistic Model (DDPM), which is trained to generate refined full-resolution fields. The DDPM leverages the FNO-predicted coarse field as a physics-informed prior, enabling accurate recovery of small-scale structures and realistic ocean dynamics. The UNET-conditioned DDPM model replaces the FNO with an UNET.}
  \label{figschematic2}
\end{figure}

\section{Results}
\label{sec:results}
In this section, we demonstrate the performance of our proposed DDPM+FNO and DDPM+UNET models as compared to traditional UNET and FNO models in terms of statistical metrics such as RMSE, CC, SSIM, as well as physics-based metrics such as the Fourier spectrum of the reconstructed variables, kinetic energy, strain rate, and relative vorticity. It must be noted that throughout this section, the reconstruction of the full-resolution fields has been generated \textbf{without an explicit need} for a background state (or generally an ensemble of background states) typically provided by a forward dynamical model at great computational cost. 

\subsection{\textbf{Performance on System 1}}
\label{sec:results_sys1}

\begin{figure}
  \centering
  \begin{overpic}[width=0.28\linewidth]{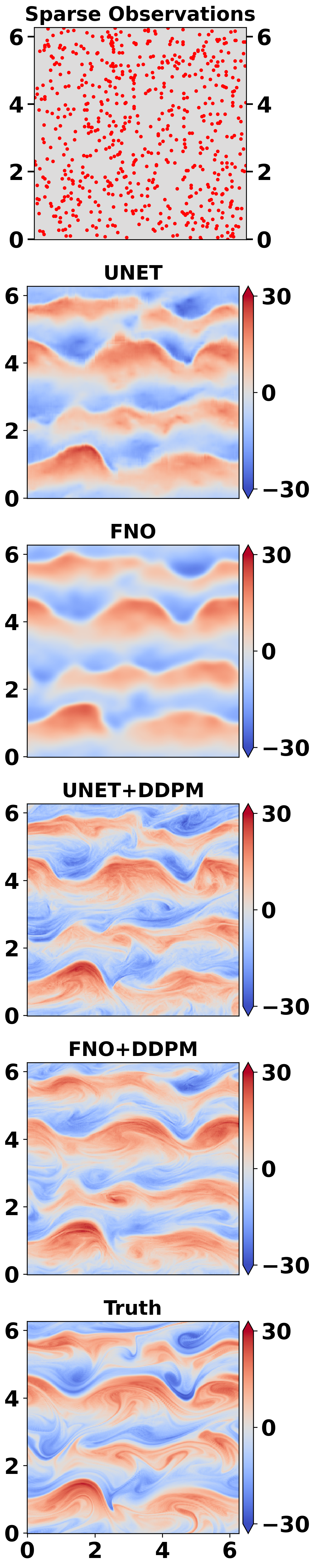}
    \put(2,100){\fontsize{12}{16}\selectfont \textbf{(a)}}
  \end{overpic}
  \hspace{0.01\linewidth}
  \begin{overpic}[width=0.28\linewidth]{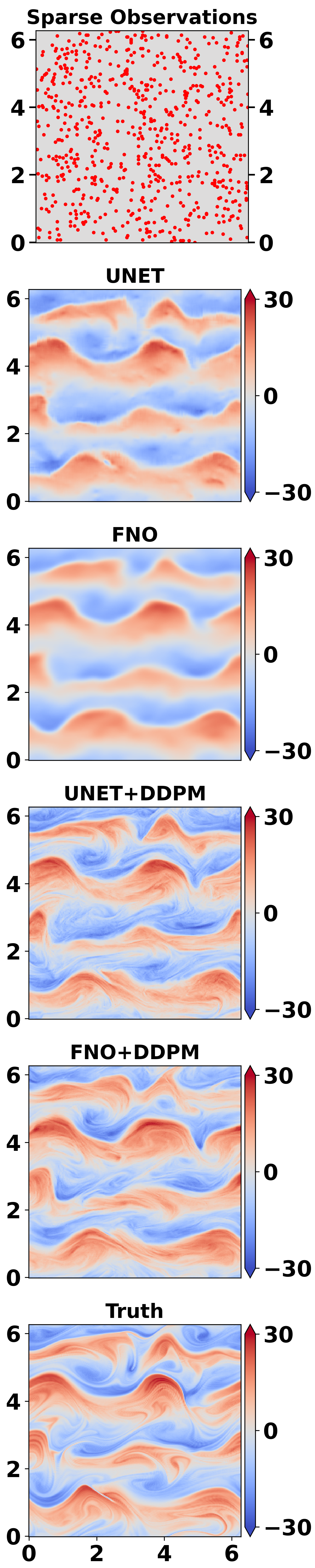}
    \put(2,100){\fontsize{12}{16}\selectfont \textbf{(b)}}
  \end{overpic}
  \hspace{0.01\linewidth}
  \begin{overpic}[width=0.28\linewidth]{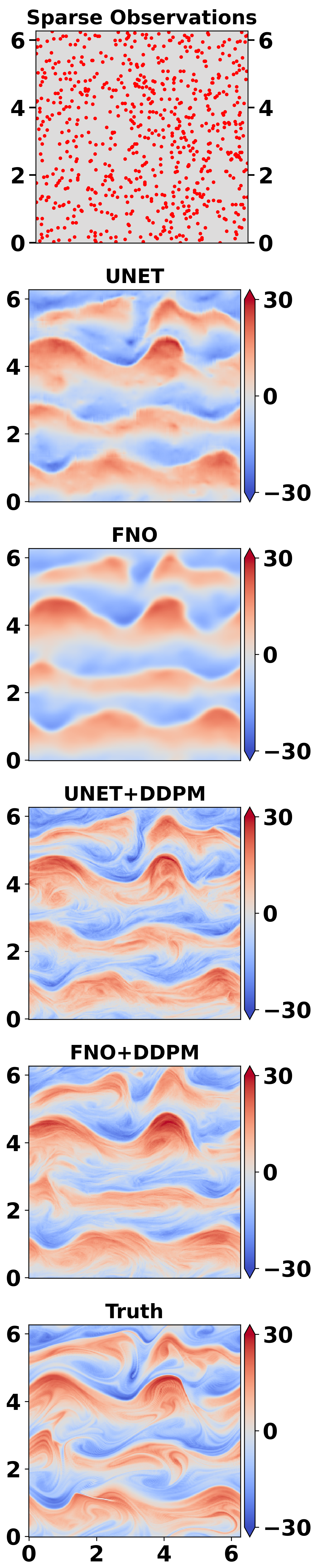}
    \put(2,100){\fontsize{12}{16}\selectfont \textbf{(c)}}
  \end{overpic}

  \caption{Vorticity reconstructions under $99\%$ sparsity. Each column shows a time snapshot; the top row shows $1\%$ observed points (red). UNET and FNO capture large-scale flow but miss fine structures. Conditional DDPM models, especially FNO+DDPM, recover sharper, more realistic small-scale features.}
  \label{turb_snapshots}
\end{figure}

Figure~\ref{turb_snapshots} presents three snapshots of the vorticity field at distinct time steps, comparing the reconstruction performance of various models under extreme observational sparsity ($99\%$). In each snapshot, the spatial locations of the observations are different, reflecting the Lagrangian or non-stationary nature of the data. The top row shows the sparse input, where red dots mark the $1\%$ available data points at each time step. The subsequent rows display reconstructions from the UNET, FNO, UNET+DDPM, and FNO+DDPM models, followed by the corresponding ground truth.

Both UNET and FNO successfully capture the large-scale structures of the vorticity field and produce smooth reconstructions; however, they fail to reproduce fine-scale features and high-wavenumber dynamics. This shortcoming is attributed to the spectral bias inherent in conventional convolutional networks and neural operators, which tend to under/over-represent high-wavenumber components~\cite{oommen2024diffusion}. As derived in Chattopadhyay et al.~\cite{chattopadhyay2023long,chattopadhyay2024oceannet} and further discussed in Lai et al.~\cite{lai2023mlclimate}, this bias is particularly problematic in climate modeling, where it can lead to instabilities and degraded performance in data-sparse regimes. Additionally, as noted in~\cite{neha2024unetreview}, the downsampling operations in UNET result in the loss of fine spatial details, which skip connections cannot fully recover, thus compromising the reconstruction of small-scale structures.

In contrast, the conditional DDPM models (UNET+DDPM and FNO+DDPM) exhibit markedly improved recovery of fine-scale vorticity patterns and sharper filaments. Among them, FNO+DDPM produces outputs with textures most closely resembling the ground truth, demonstrating a superior ability to reconstruct multi-scale turbulent features under conditions of extreme and time-varying sparsity.

\begin{figure}
  \centering
  \includegraphics[width=1\textwidth]{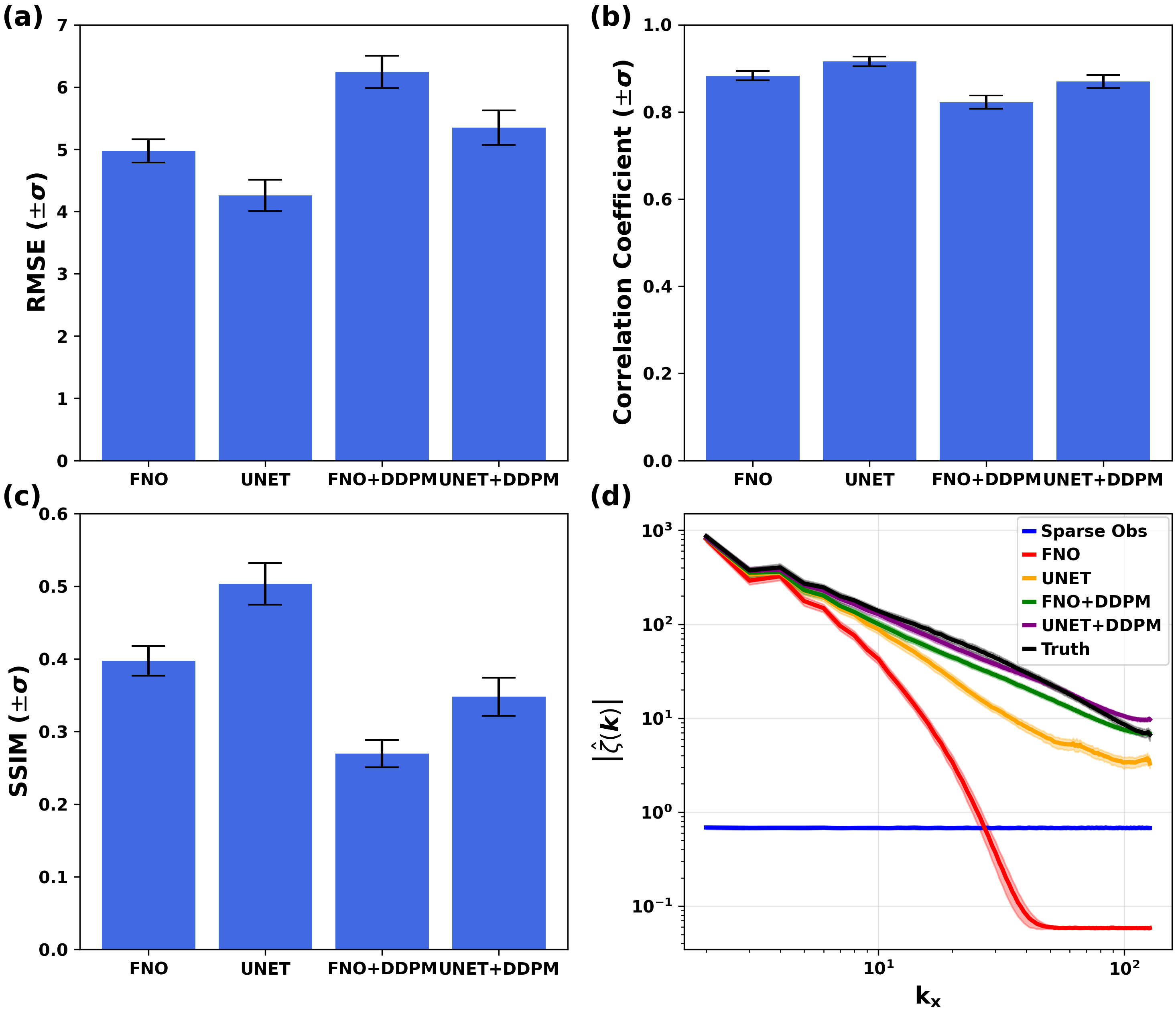}
     \caption{Reconstruction performance across 100 test samples using (a) RMSE, (b) correlation coefficient (CC), (c) SSIM, and (d) Fourier spectrum. UNET and FNO show favorable RMSE, CC, and SSIM due to large-scale alignment but miss fine-scale features. Conditional DDPM models, despite slightly higher RMSE, better preserve high-wavenumber energy in (d), indicating improved recovery of small-scale turbulent structures.}
  \label{vorticity_metrics}
\end{figure}

Figure~\ref{vorticity_metrics} compares the reconstruction performance of all models using four evaluation metrics computed over more than 100 test samples. Figures~\ref{vorticity_metrics}(a), (b), and (c) report the RMSE, CC, and SSIM, respectively, while panel (d) presents the Fourier spectrum of the reconstructed vorticity fields. While RMSE, CC, and SSIM are widely used in image reconstruction tasks, they are often inadequate and even misleading for evaluating the physical fidelity of turbulent flows. These metrics primarily emphasize agreement at large spatial scales and pointwise similarity while systematically under-representing errors in small-scale structures that are crucial for capturing the true dynamics of turbulent systems. The error bar in each of these metrics is computed across $100$ test samples outside the training dataset. 

For instance, the UNET model achieves the lowest average RMSE, as shown in Figure~\ref{vorticity_metrics}(a), due to its ability to match the broad, smooth patterns of the ground truth field. However, as evident in Figure~\ref{turb_snapshots}, its outputs are overly diffused and lack small-scale detail. RMSE, being a pointwise error metric, is insensitive to structural realism and physical coherence. Similarly, while the correlation coefficient and SSIM (Figure~\ref{vorticity_metrics}(b) and (c) are higher for UNET and FNO, these metrics primarily reflect large-scale alignment and do not penalize the loss of high-wavenumber components. In contrast, the conditional DDPM models (UNET+DDPM and FNO+DDPM), which produce sharper, more filamented reconstructions, report slightly higher RMSE and lower SSIM—despite yielding outputs that better preserve the underlying physics.

To address these limitations, we incorporate a spectral analysis based on the Fourier spectrum (Figure~\ref{vorticity_metrics}(d)), which more appropriately captures the distribution of energy across spatial scales. The spectra for UNET and FNO exhibit a pronounced drop in power at high wavenumbers, indicating a failure to reconstruct fine-scale vorticity features, a known manifestation of spectral bias in conventional neural networks. In contrast, the spectra of FNO+DDPM and UNET+DDPM closely follow those of the ground truth, particularly in the high-wavenumber regime, demonstrating improved recovery of small-scale structures. These results highlight the importance of evaluating reconstructed turbulent fields using physics-informed diagnostics, such as spectral energy content, which are sensitive to the multi-scale nature of turbulence and are not captured by traditional statistical metrics.

\begin{figure}
  \centering
  \includegraphics[width=0.68\textwidth]{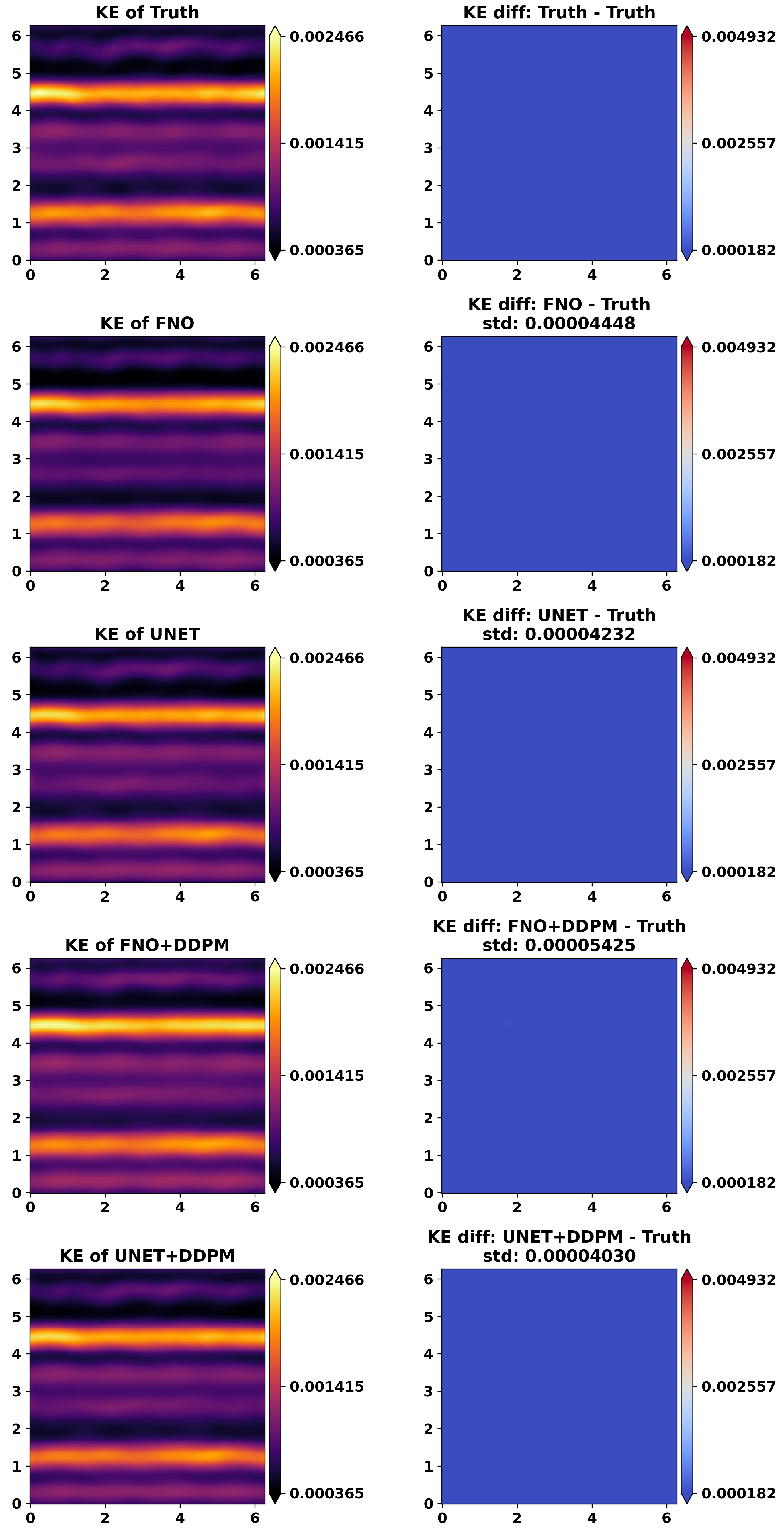}
     \caption{Average kinetic energy (KE) fields and absolute error maps over 3000 samples. Left: temporally averaged KE fields for the ground truth and each model over $3000$ samples. Right: pointwise absolute differences from the ground truth with standard deviation. All models, especially the DDPM-enhanced ones, accurately capture large-scale KE despite $99\%$ sparsity, demonstrating preservation of bulk flow physics beyond mere interpolation.}
  \label{KE_vorticity}
\end{figure}

Figure~\ref{KE_vorticity} presents the average kinetic energy (KE) fields computed over $3000$ samples for the ground truth and all reconstructed models, along with corresponding pointwise absolute difference maps. The left column displays the spatially averaged KE fields for each model, while the right column shows their deviation from the ground truth, quantified by pointwise absolute error and associated standard deviation values. Overall, all models exhibit similar performance in reconstructing the bulk kinetic energy of the vorticity field. This is expected, as spatially averaged KE is dominated by contributions from large-scale structures and is relatively insensitive to fine-scale fluctuations. The primary motivation for presenting this figure is to highlight that, despite the extreme observational sparsity ($99\%$) and dynamically changing observation locations across time, all models—particularly the DDPM-enhanced variants—successfully reconstruct the vorticity field while preserving accurate large-scale energy content. This demonstrates that the learned models are not only interpolating observations but are capturing the essential bulk physics of the system, as reflected in their ability to reproduce the domain-averaged KE with high fidelity.

\subsection{\textbf{Performance on System 2}}

\begin{figure}
  \centering

  \begin{overpic}[width=0.30\linewidth]{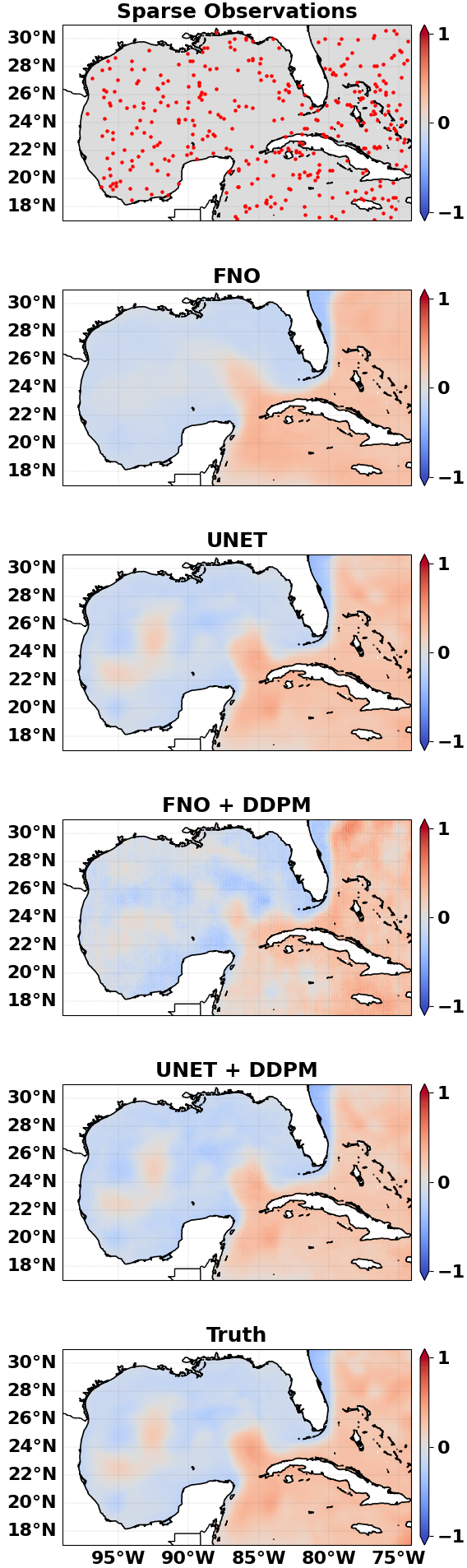}
    \put(2,100){\fontsize{12}{16}\selectfont \textbf{(a)}}
  \end{overpic}
  \hspace{0.01\linewidth}
  \begin{overpic}[width=0.30\linewidth]{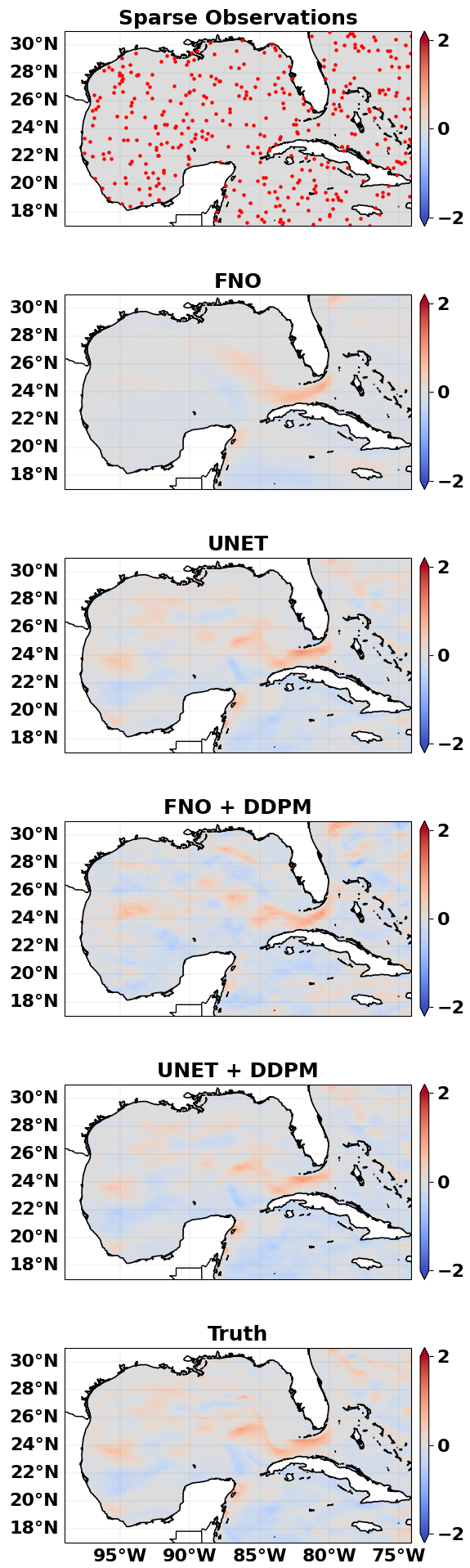}
    \put(2,100){\fontsize{12}{16}\selectfont \textbf{(b)}}
  \end{overpic}
  \hspace{0.01\linewidth}
  \begin{overpic}[width=0.30\linewidth]{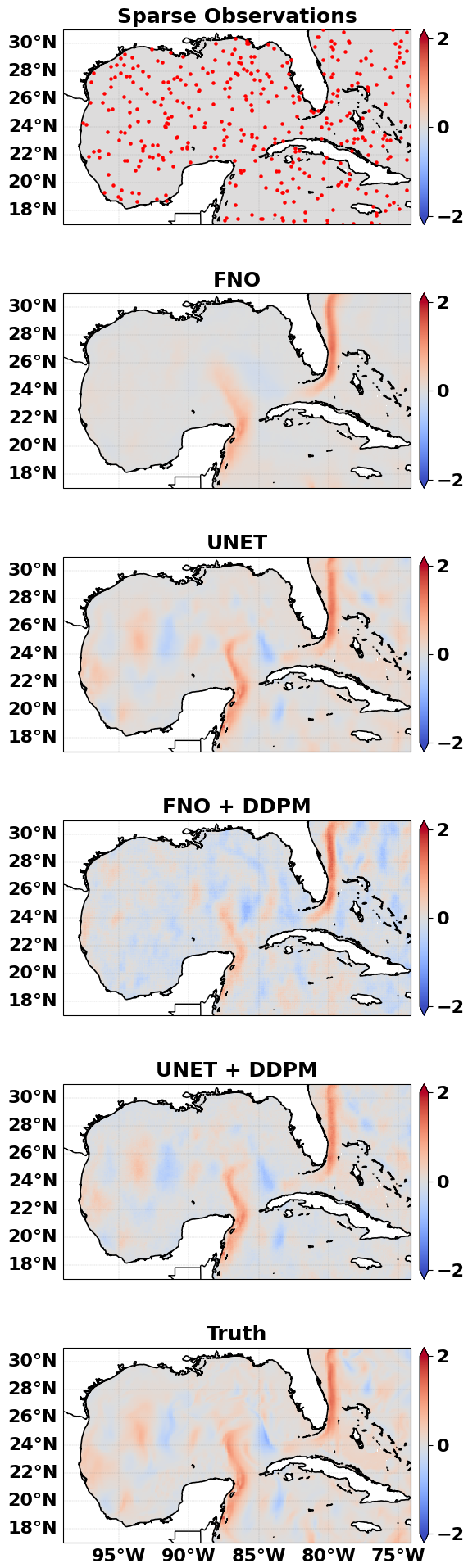}
    \put(2,100){\fontsize{12}{16}\selectfont \textbf{(c)}}
  \end{overpic}

  \caption{Reconstructed fields over the Gulf of Mexico on January 1, 2013: (a) SSH, (b) SSU, and (c) SSV. Top rows show $1\%$ observed input (red dots). UNET and FNO produce oversmoothed outputs lacking fine structure, while conditional DDPM models better capture spatial variability, especially in the more dynamic SSU and SSV fields.}
  \label{GOM_snapshots}
\end{figure}

Figure~\ref{GOM_snapshots} displays reconstructed fields of SSH in (a), SSU in (b), and SSV in (c) over the Gulf of Mexico on January 1, 2013. The top row of each panel shows the $99\%$-sparse input, where red dots indicate the $1\%$ available observations. The base models (UNET and FNO) fail to reconstruct fine-scale flow structures and instead produce overly smooth outputs with limited spatial variability. In contrast, the conditional DDPM models (UNET+DDPM and FNO+DDPM) exhibit substantial improvement, capturing both spatial variability and fine-scale features more faithfully. This improvement is particularly notable in the SSU and SSV fields, which are inherently more dynamic and sensitive to transient flow features than SSH.

\begin{figure}
  \centering
  \includegraphics[width=1\textwidth]{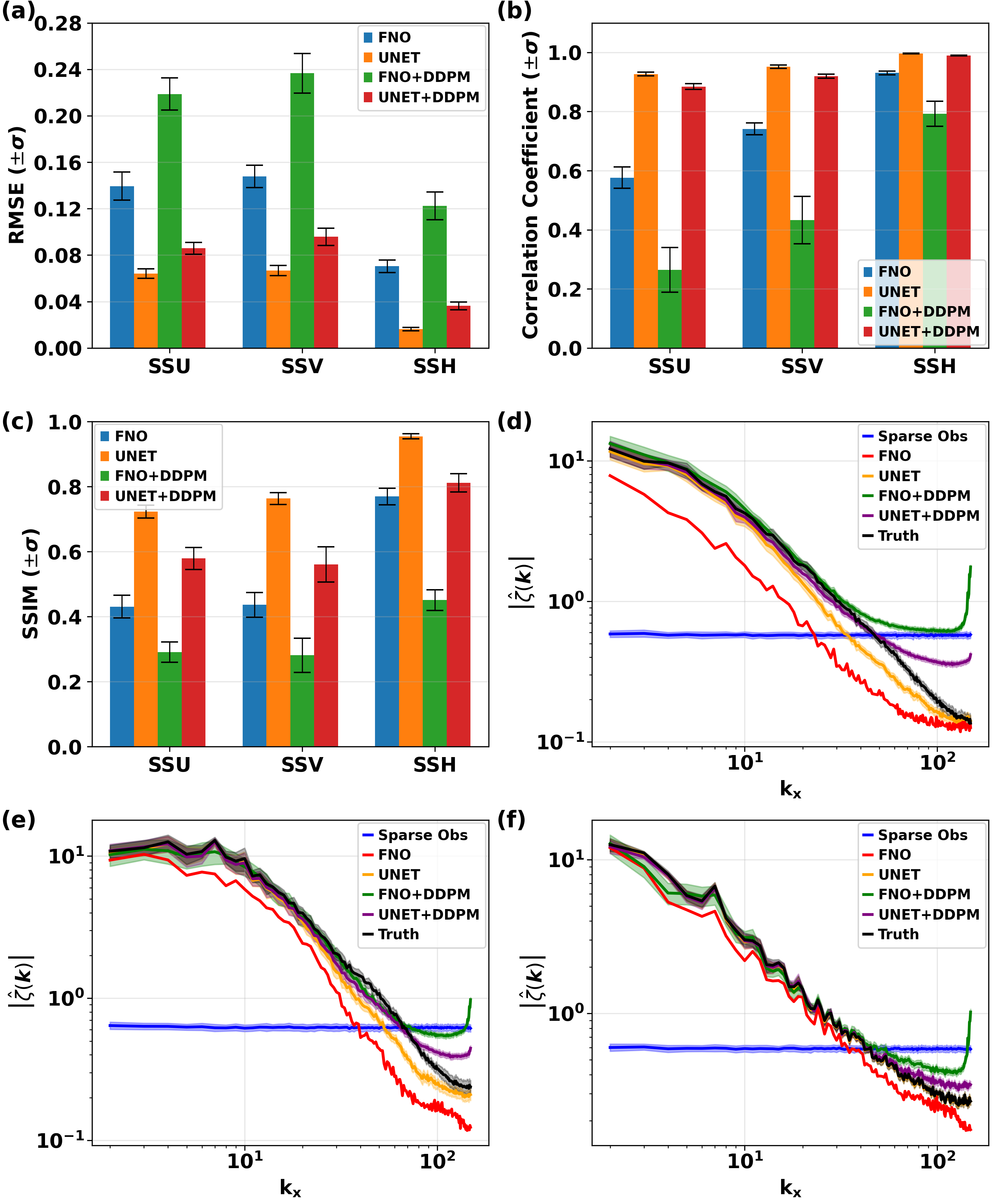}
  \caption{Quantitative comparison over $50$ test samples: (a) RMSE, (b) CC, (c) SSIM, and mean Fourier spectra of (d) SSU, (e) SSV, and (f) SSH. All models capture the SSH spectrum well due to its large-scale nature, though conditional DDPM models perform best at high wavenumbers. For SSU and SSV, FNO and UNET fail to recover a wide range of spatial scales, while conditional DDPM models better match the ground truth spectra, especially at intermediate wavenumbers, despite some bias at the highest wavenumbers.}
  \label{GOM_metrics}
\end{figure}

Figure~\ref{GOM_metrics} summarizes the quantitative comparison of model performance over 50 test samples. Figure~\ref{GOM_metrics}(a) shows RMSE, (b) presents the average CC, and (c) shows the structural similarity index (SSIM) over $50$ test samples. Figure~\ref{GOM_metrics}(d), (e), and (f) display the mean Fourier spectra of SSU, SSV, and SSH, respectively. As shown in panel (f), the SSH spectrum is well captured by all the models (although the conditional DDPM models outperform the base models). This is because SSH has predominantly large-scale features, which are well captured by the base models. However, both the base models, FNO and UNET, underestimate the SSH dynamics at higher wavenumbers. In contrast, the spectra for SSU and SSV in panels (d) and (e) reveal that the base models fail to recover a broad range of spatial scales—FNO failing to capture even the lowest wavenumbers while UNET captures about half the wavenumbers. The performance of FNO+DDMP or UNET+DDPM is best seen in these panels, where the spectrum of both SSU and SSV is well captured by the generative models, albeit significant spectral bias can be seen in the highest wavenumbers.

\begin{figure}
  \centering
  \includegraphics[width=0.8\textwidth]{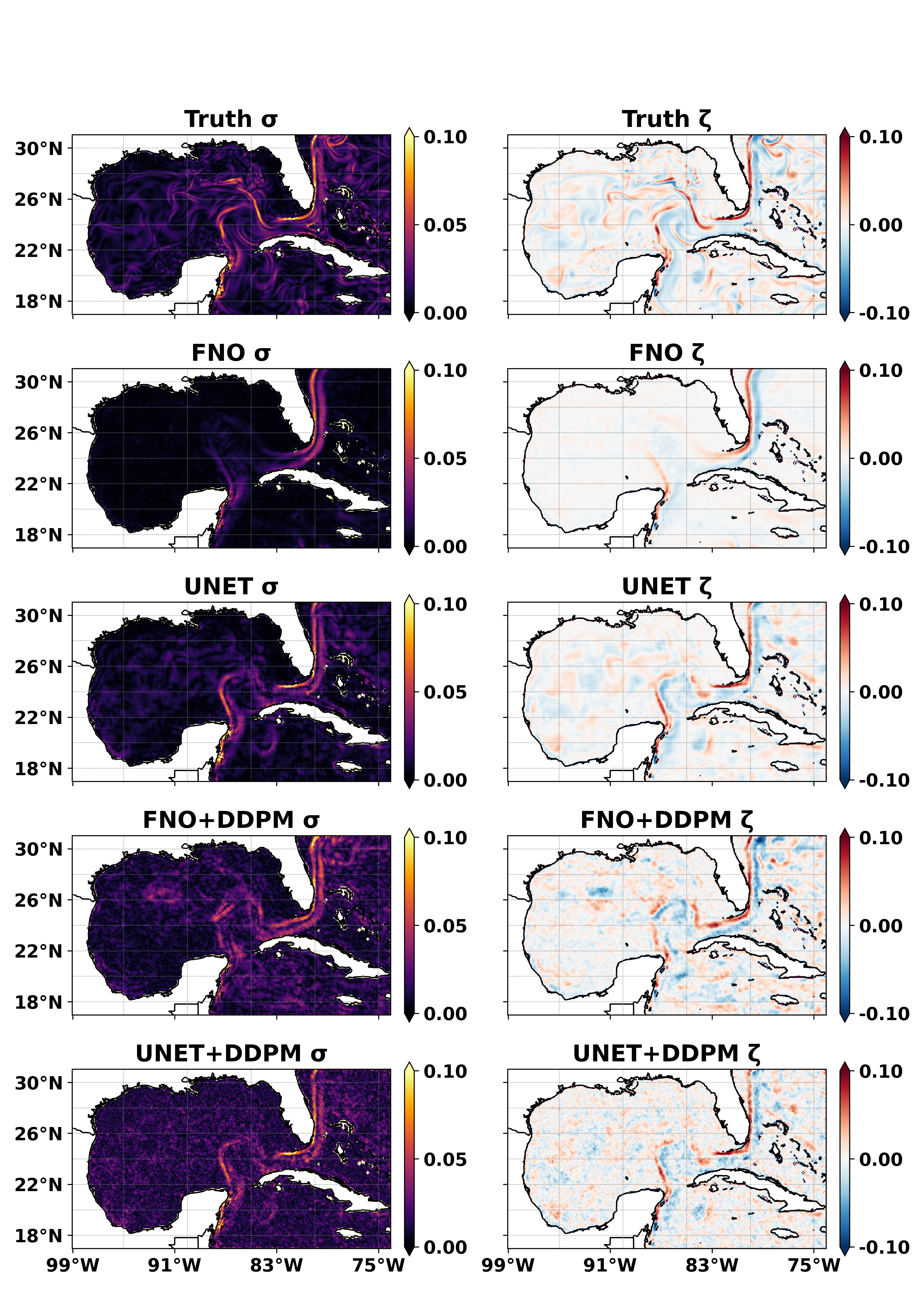}
  \caption{Derived diagnostics: (a) strain rate \( \sigma \) and (b) relative vorticity\( \zeta \)—computed from reconstructed SSU and SSV fields. Ground truth shows sharp eddy and filament structures near the Loop Current. FNO captures smooth patterns but lacks detail; UNET delineates eddies better but introduces artifacts. Conditional DDPM models improve gradient localization, with UNET+DDPM most closely matching the ground truth in both structure and intensity.}
  \label{GOM_strainzeta}
\end{figure}

Figure~\ref{GOM_strainzeta} evaluates the physical consistency of the reconstructed velocity fields by showing derived diagnostics, e.g., strain rate ($\sigma$) and relative vorticity ($\zeta$) computed from SSU and SSV. The diagnostic expressions follow Martin et al.~\cite{martin2024genda}, and are given by:
\begin{equation}
\sigma = \sqrt{\sigma_n^2 + \sigma_s^2} = 
\sqrt{\left( \frac{\partial u}{\partial x} - \frac{\partial v}{\partial y} \right)^2 + 
\left( \frac{\partial v}{\partial x} + \frac{\partial u}{\partial y} \right)^2}.
\label{eq:sigma}
\end{equation}

\begin{equation}
\zeta = \frac{\partial v}{\partial x} - \frac{\partial u}{\partial y}.
\label{eq:zeta}
\end{equation}

The ground truth fields exhibit sharp mesoscale eddies and filamentary structures, especially near the Loop Current. FNO captures smooth patterns but lacks small-scale resolution. UNET introduces some artifacts but delineates eddy boundaries better than FNO. FNO+DDPM shows improved localization of gradients and finer structures, though it introduces mild noise. UNET+DDPM most closely resembles the ground truth in both spatial pattern and intensity, showing that DDPM correction significantly enhances the physical consistency of reconstructed fields across multiple scales.

\subsection{\textbf{Performance on System 3}}

\begin{figure}
  \centering

  \begin{overpic}[width=0.30\linewidth]{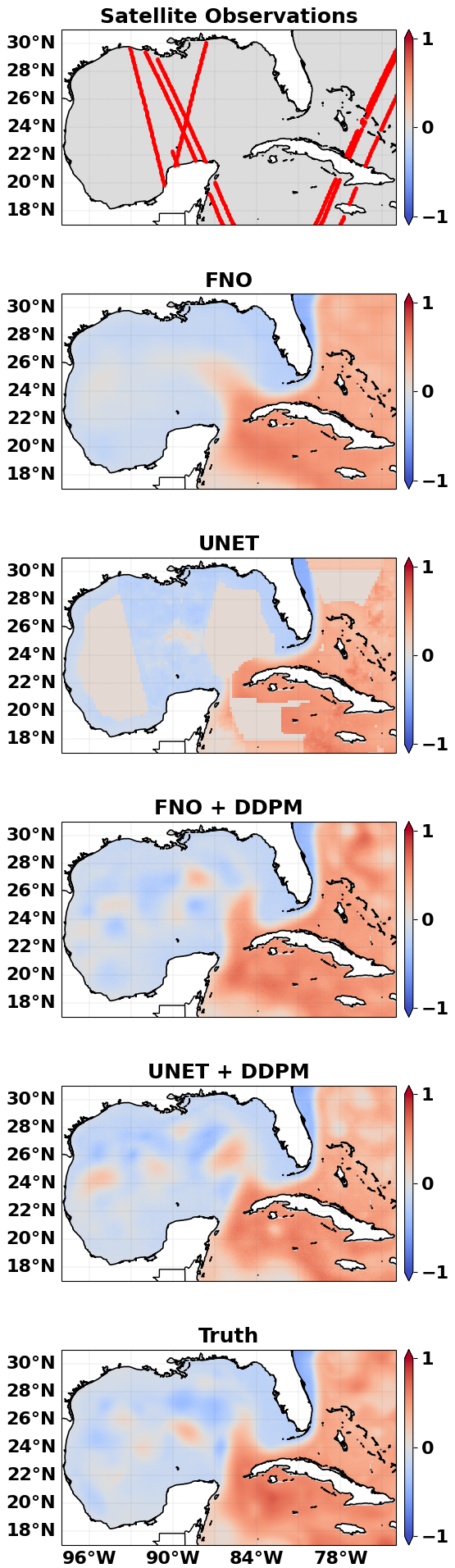}
    \put(2,100){\fontsize{12}{16}\selectfont \textbf{(a)}}
  \end{overpic}
  \hspace{0.01\linewidth}
  \begin{overpic}[width=0.30\linewidth]{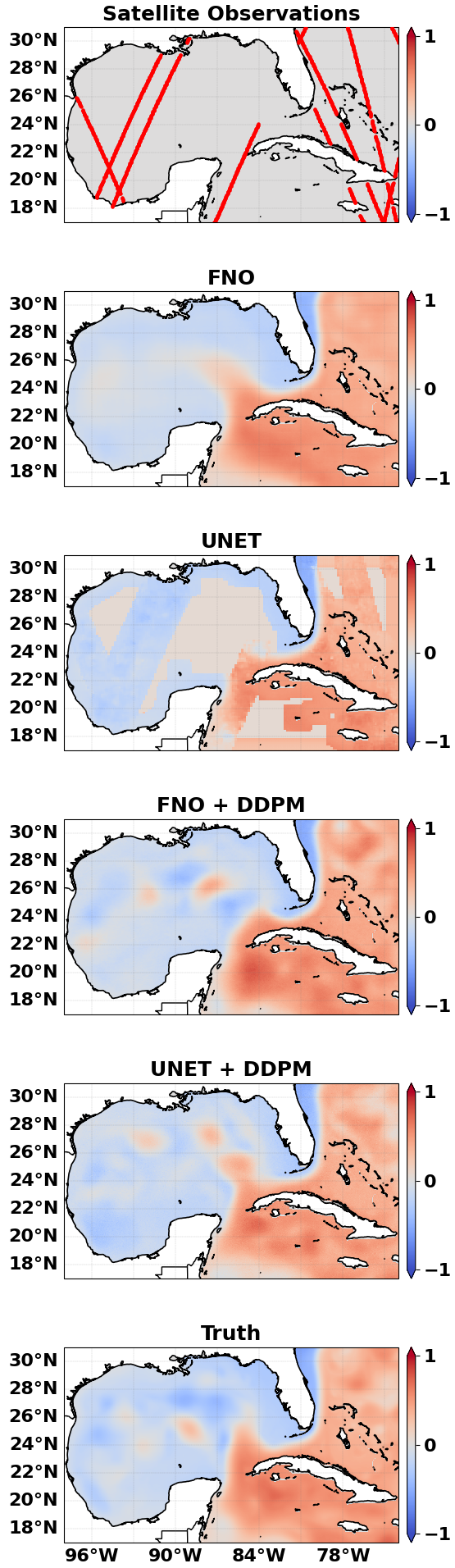}
    \put(2,100){\fontsize{12}{16}\selectfont \textbf{(b)}}
  \end{overpic}
  \hspace{0.01\linewidth}
  \begin{overpic}[width=0.30\linewidth]{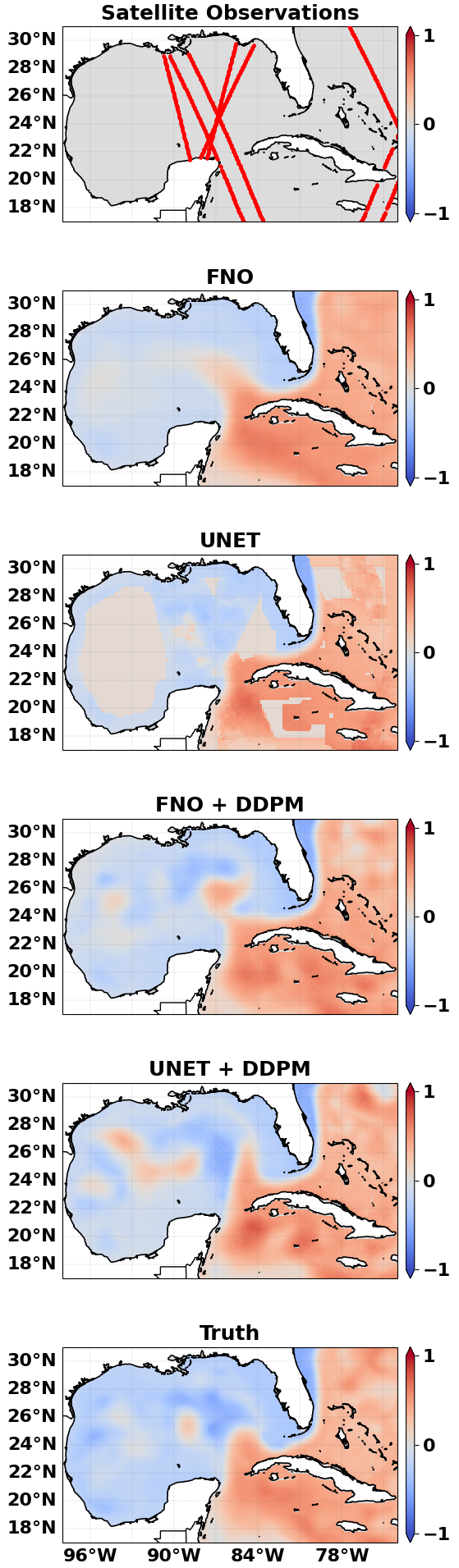}
    \put(2,100){\fontsize{12}{16}\selectfont \textbf{(c)}}
  \end{overpic}

  \caption{Three snapshots of satellite-observed sea surface height (SSH) over the Gulf of Mexico under extreme sparsity (99.7\%). UNET fails to generalize and overfits observed points; FNO improves continuity but oversmooths. Conditional DDPM models better recover fine-scale features, showing greater robustness to sparse, non-stationary observations.}
  \label{satt_snapshots}
\end{figure}

Figure~\ref{satt_snapshots} shows three representative snapshots of satellite-observed sea surface height (SSH) over the Gulf of Mexico on distinct days. Due to the extreme sparsity ($\sim$99.7\%) and temporally varying observation locations, this constitutes one of the most challenging reconstruction settings considered. While UNET performs well on synthetic and reanalysis data, it fails to generalize to real-world observations, producing incoherent reconstructions that largely overfit to observed points while collapsing in unobserved regions. FNO yields better spatial continuity but suffers from excessive smoothing, particularly at smaller scales. In contrast, the DDPM-conditioned models (UNET+DDPM and FNO+DDPM) reconstruct fine-scale SSH features more consistently across the domain, indicating their superior capacity to model spatial complexity under severe data sparsity.

\begin{figure}
  \centering
  \includegraphics[width=1\textwidth]{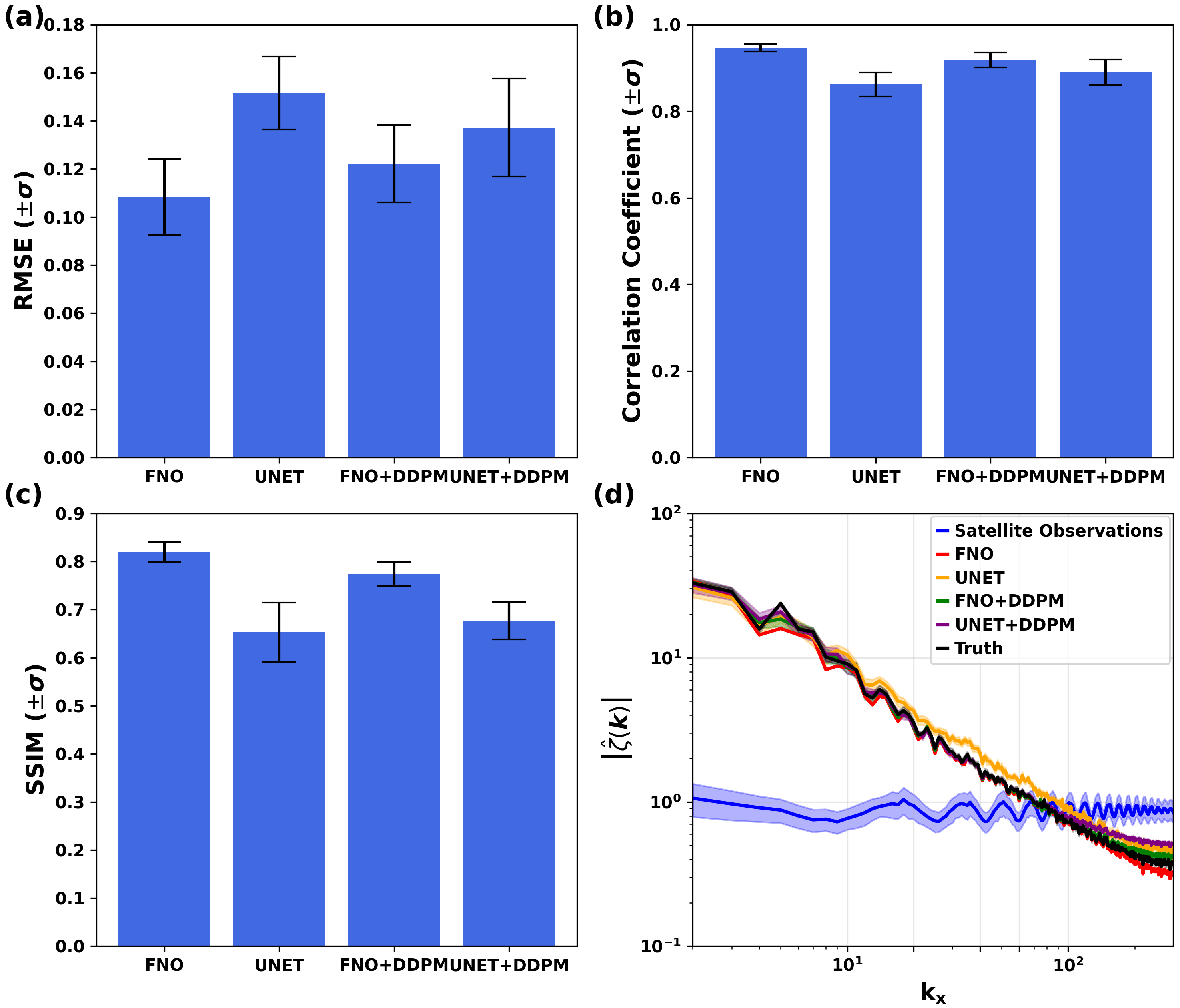}
     \caption{Reconstruction performance averaged over $50$ days: (a) RMSE, (b) correlation coefficient, (c) SSIM, and (d) FFT power spectrum. UNET performs worst across all metrics. FNO achieves lower RMSE but lacks physical fidelity. Conditional DDPM models better match the ground truth spectrum and preserve small-scale dynamics, spatial variability, and structural integrity—key for oceanographic accuracy.}
  \label{satt_metrics}
\end{figure}

Figure~\ref{satt_metrics} reports the averaged reconstruction performance over 50 days. Panel (a) presents the RMSE, panel (b) the correlation coefficient (CC), panel (c) the SSIM, and panel (d) the FFT power spectrum. UNET exhibits the highest RMSE and lowest SSIM and CC, confirming its poor generalization to sparse observational input. Conditioning with DDPM markedly improves its performance in all metrics. Although FNO achieves the lowest RMSE, this pointwise metric alone fails to capture the physical plausibility of the reconstruction. The power spectra in panel (d) show that the conditional DDPM models closely match the ground truth across wavenumbers, particularly retaining energy at higher wavenumbers that represent small-scale dynamics. CC and SSIM results further support that these models better preserve spatiotemporal variability and structural integrity. This is especially critical for oceanographic applications, where accurately resolving fronts, eddies, and gradient structures is essential for analysis and forecasting.


\vspace*{-2 cm}

\setlength{\intextsep}{2pt}
\setlength{\textfloatsep}{2pt}

\setlength{\intextsep}{2pt}
\setlength{\textfloatsep}{2pt}

\clearpage
\section{Conclusion and Discussion}

In this study, we introduced a novel generative modeling framework for reconstructing oceanic surface fields from highly sparse and non-stationary Lagrangian satellite observations. By conditioning denoising diffusion probabilistic models (DDPMs) on Fourier Neural Operator (FNO) predictions, we demonstrated significant improvements in recovering fine-scale flow structures, spectral consistency, and physical diagnostics such as vorticity and strain, compared to existing baselines. These advances are particularly important in the context of climate and ocean forecasting, where observations are often limited, noisy, and spatially inconsistent.

Unlike traditional data assimilation (DA) schemes, which require access to computationally expensive forward dynamical models like ROMS or MOM6, our approach is purely data-driven and bypasses the need for explicit numerical solvers. In variational or ensemble-based DA systems, the assimilation step relies on tangent-linear and adjoint models or multiple ensemble integrations, which are costly and prone to numerical instability, especially under extreme data sparsity. In contrast, our generative model provides an efficient one-shot reconstruction pipeline that can handle high sparsity and time-varying observation geometries without requiring any knowledge of governing equations or solver access. This positions our framework as a strong alternative for rapid reconstruction in operational systems or in scenarios where computational resources are constrained.

Future directions involve systematically validating our reconstructions by using them as initial conditions in forecast models. One approach is to initialize OceanNet, a data-driven forecasting model introduced by Chattopadhyay et al.~\cite{chattopadhyay2024oceannet} and extended by Jiménez et al.~\cite{lupin2025simultaneous}, with our DDPM-based reconstructions. Because OceanNet has been shown to successfully predict ocean dynamics over extended time horizons, its error growth when initialized with different reconstruction methods could serve as a proxy for the accuracy of our framework. Similarly, initializing a dynamical model such as ROMS with our reconstructions can provide physical validation through forecast skill metrics. A comparison of forecast horizon and energy spectra of the ROMS outputs across different reconstruction initializations would yield insights into which method best captures the true underlying dynamics.

Furthermore, our generative framework can be extended to perform global ocean reconstructions. The scalability of FNOs and convolutional backbones, along with conditional DDPMs, makes this pipeline naturally extensible to global nadir-based observations such as satellite altimetry data from Jason missions, as well as newly available swath-based Synthetic Aperture Radar (SAR) data from the Surface Water and Ocean Topography (SWOT) mission. The resulting full-resolution surface state estimates can be used to initialize coupled climate models. One promising direction is to integrate our global reconstructions with LUCIE, a data-driven climate emulator developed by Guan et al.~\cite{guan2024lucie}, to enable coupled atmosphere-ocean simulations that are both high-resolution and computationally tractable. Such modeling approaches could bring us closer to enabling scalable Earth system modeling that is rooted in observations, not just reanalysis products.

Our methodology also opens the door to reconstructing subsurface ocean fields, such as thermocline depth, salinity, and deep currents. By training on multi-modal observations from Argo floats, deep moorings, and altimetry-derived surface fields, we can potentially learn inverse mappings from sparse surface data to volumetric fields. This could dramatically expand the observational footprint in otherwise inaccessible ocean layers.

Finally, a significantly longer-term goal is to shift from using reanalysis fields as training labels to directly learning from raw satellite and in situ observations. Such a fully data-driven reconstruction pipeline would eliminate the dependency on models like HYCOM or CMEMS, which themselves embed strong dynamical priors. Instead, our generative framework would learn physical consistency directly from the data, unlocking the possibility of more flexible, observationally grounded ocean modeling across spatiotemporal scales and regimes.

In summary, our results suggest that physics-aware generative modeling, conditioned on learned operators, offers a scalable, accurate, and computationally efficient alternative to traditional ocean data assimilation systems, particularly under conditions of extreme sparsity and observational noise. With proper integration into forecasting and coupled modeling pipelines, this approach holds significant promise for advancing ocean monitoring, prediction, and understanding.

\section{Acknowledgement}
AC and RH designed the research. NA and LJ conducted the research and developed the computational models with equal contributions. TW provided the satellite observations. NA and AC wrote the manuscript. All authors analyzed the results and edited the manuscript. AC and LJ were supported by the National Science Foundation (grant no. 2425667). Computational resources were provided by NSF ACCESS MTH240019, MTH250006 and NCAR CISL UCSC0008, and UCSC0009.

\section{Data and Code}
The computational models used in this study are publicly available in \url{https://github.com/TACS-UCSC/GenDA-Lagrangian}.

\section{References}
\bibliographystyle{iopart-num}
\bibliography{references}

\end{document}